\documentclass[conference]{IEEEtran}
\IEEEoverridecommandlockouts
\usepackage{bbm}
\usepackage{graphicx}
\usepackage{xcolor}
\usepackage{float}   
\usepackage[normalem]{ulem}
\usepackage{dsfont}
\usepackage{textcomp}
\usepackage{stfloats}
\usepackage{url}
\usepackage{verbatim}
\usepackage{algorithmic}
\usepackage{array}
\usepackage{booktabs}
\usepackage{pifont}
\usepackage{comment}
\usepackage{algorithm}
\usepackage{algorithmic}
\usepackage[acronym]{glossaries}
\newacronym{ai}{AI}{Artificial Intelligence}
\newacronym{aigc}{AIGC}{Artificial Intelligence Generated Content}
\newacronym{ack}{ACK}{Acknowledgement}

\newacronym{aoi}{AoI}{Age of Information}

\newacronym{awgn}{AWGN}{Additive White Gaussian Noise}

\newacronym{bpp}{bpp}{bits per pixel}

\newacronym{cca}{CCA}{Clear Channel Assessment}
\newacronym{csma}{CSMA}{Carrier Sense Multiple Access}
\newacronym{cnn}{CNN}{Convolution Neural Network}
\newacronym{dl}{DL}{Deep Learning}
\newacronym{dnn}{DNN}{Deep Neural Network}
\newacronym{dram}{DRAM}{Dynamic Random Access Memory}

\newacronym{es}{ES}{edge server}

\newacronym{fp}{FP}{False Positive}
\newacronym{fn}{FN}{False Negative}
\newacronym{fm}{FM}{Foundation Model}
\newacronym{fid}{FID}{Fr\'{e}chet Inception Distance}
\newacronym{fifo}{FIFO}{First In First Out}

\newacronym{gan}{GAN}{Generative Adversarial Network}
\newacronym{gai}{G-AI}{Generative AI}
\newacronym{giot}{GIoT}{Generative Internet of Things}
\newacronym{gpt}{GPT}{Generative Pre-trained Transformer}

\newacronym{hific}{HiFiC}{High Fidelity Compression}

\newacronym{iot}{IoT}{Internet of Things}
\newacronym{id}{ID}{Identity}
\newacronym{idwu}{IDWu}{Identity-based Wake-up}
\newacronym{sifi}{SiFi}{Significance and Fidelity}
\newacronym{tinyair}{TinyAirNet}{Tiny Neural Network transmission over the Air}

\newacronym{jscc}{JSCC}{Joint Source Channel Coding}

\newacronym{kpi}{KPI}{Key Performance Indicator}

\newacronym{llm}{LLM}{Large Language Model}
\newacronym{lpips}{LPIPS}{Learned Perceptual Image Patch Similarity}
\newacronym{lpwan}{LP-WANs}{Low-Power Wide-Area Networks}
\newacronym{mac}{MAC}{Medium Access Control}
\newacronym{ml}{ML}{Machine Learning}
\newacronym{ofdma}{OFDMA}{Orthogonal Frequency Domain Multiple Access}

\newacronym{vae}{VAE}{Variational Autoencoder}

\newacronym{mcstep}{MC step}{Monte Carlo step}
\newacronym{mcmc}{MCMC}{Markov Chain Monte Carlo}

\newacronym{mcu}{MCUs}{Micro Controller Unit}

\newacronym{muac}{MUAC}{Multiply-Accumulate}
\newacronym{madd}{mADD}{multiplication and addition}

\newacronym{ook}{OOK}{On-Off Keying}
\newacronym{pmf}{PMF}{Probability Mass Function}
\newacronym{qnn}{QNN}{Quantized Neural Network}

\newacronym{semdas}{SEMDAS}{SEMantic DAta Soucing}
\newacronym{sram}{SRAM}{Static Random Access Memory}

\newacronym{rr}{RR (w/PNG)}{Round-robin scheduling}
\newacronym{tp}{TP}{True Positive}
\newacronym{tn}{TN}{True Negative}

\newacronym{wsn}{WSNs}{Wireless Sensor Networks}

\newacronym{uav}{UAV}{Unmanned Aerial Vehicle}
\usepackage{amssymb,amsmath,amsfonts,amsthm,bm,bbm,mathtools,cuted,bbold}
\usepackage[normalem]{ulem}

\usepackage[cmintegrals]{newtxmath}

\usepackage{multirow}

\usepackage[font=small]{caption}
\usepackage[font=footnotesize]{subcaption}
\usepackage{tikz}
\usetikzlibrary{plotmarks,patterns,decorations.pathreplacing,backgrounds,calc,arrows,arrows.meta,spy,matrix}
\usepackage{tikzscale}

\usepackage{pgfplots}
\usepgfplotslibrary{colorbrewer, groupplots, patchplots}
\pgfplotsset{
    compat=newest,
    legend style={font=\footnotesize, fill opacity=0.7,  draw opacity=1, text opacity=1, draw=white!15!black, legend cell align=left, align=left}, 
    width=0.8\columnwidth, 
    scale only axis,
    height=4cm,
    yminorticks=false,
    xminorticks=false,
    label style={font=\small},
    title style={font=\small},
    tick align=outside,
    tick pos=left,
    tick style={color=black},
    tick label style={font=\footnotesize},
    grid style={line width=.1pt, draw=gray!20},
    major grid style={line width=.1pt,draw=gray!20},
    plot coordinates/math parser=false 
}
\newlength\figureheight
\newlength\figurewidth

\usepackage{multirow}
\usepackage{tablefootnote}
\usepackage{booktabs}
\usepackage{tabularx}

\def\BibTeX{{\rm B\kern-.05em{\sc i\kern-.025em b}\kern-.08em T\kern-.1667em\lower.7ex\hbox{E}\kern-.125emX}}

\newcommand{\argmin}{\mathop{\rm arg~min}\limits}
\newcommand{\probP}{\text{I\kern-0.15em P}}

\definecolor{amaranth}{rgb}{0.9, 0.17, 0.31}

\begin{document}

\title{
    EcoPull: Sustainable IoT Image Retrieval \\ Empowered by TinyML Models
    \thanks{
        \textsuperscript{\text{*}}Mathias T., Victor C., and Junya S. contributed equally to this work. This work was partly supported by the Villum Investigator Grant ``WATER" from the Velux Foundation, Denmark, partly by the Horizon Europe SNS ``6G-XCEL" project with Grant 101139194, and partly by the Horizon Europe SNS ``6G-GOALS'' project with grant 101139232.
    }
}

\author{
    Mathias Thorsager\textsuperscript{\text{*}},~Victor Croisfelt\textsuperscript{\text{*}},~Junya Shiraishi\textsuperscript{\text{*}},~and~Petar Popovski~\\
    Department of Electronic Systems, Aalborg University, Denmark\\
    Email: $\{$mdth,~vcr,~jush,~petarp$\}$@es.aau.dk
}

\maketitle

\begin{abstract}
    This paper introduces EcoPull, a sustainable Internet of Things (IoT) framework empowered by tiny machine learning (TinyML) models for fetching images from wireless visual sensor networks. Two types of learnable TinyML models are installed in the IoT devices: i) a \emph{behavior model} and ii) an \emph{image compressor model}. The first filters out irrelevant images for the current task, reducing unnecessary transmission and resource competition among the devices. The second allows IoT devices to communicate with the receiver via latent representations of images, reducing communication bandwidth usage. However, integrating learnable modules into IoT devices comes at the cost of increased energy consumption due to inference. The numerical results show that the proposed framework can save $ > 70\%$ energy compared to the baseline while maintaining the quality of the retrieved images at the ES.
\end{abstract}

\section{Introduction}
One of the key requirements for \gls{iot} devices in the context of 6G is sustainability~\cite{strinati20216g,sheth2020taxonomy}. The \gls{iot} devices will also be affected by the trend of using \gls{ai}/\gls{ml} to support various intelligent tasks at the edge~\cite{strinati2024goal}. However, \gls{iot} devices are challenged by resource constraints in terms of energy, computation, and memory, which hinders the implementation of learning models on these devices. First, the practical \gls{mcu} of the \gls{iot} devices only have a small amount of memory and storage, \emph{e.g.}, on-chip \gls{sram} ($<512$ KB) and flash storage ($<2$ MB)~\cite{lin2020mcunet}. Second, running \gls{ml} models require additional energy cost for \gls{iot} devices. Therefore, one must design a communication protocol considering the energy cost associated with the \gls{ml} model and the main radio interface. 

In~\cite{shiraishi2023energy}, we introduced \gls{tinyair}, an \gls{iot} image retrieval framework that reduces the overall energy consumption of the \gls{iot} devices. In \gls{tinyair}, a Tiny\gls{ml} model is transmitted by the \gls{es} to the \gls{iot} devices before they collect new images to facilitate the subsequent semantic query, by which the \gls{iot} device selectively transmits relevant images and avoids energy waste. Although this enables the suppression of images likely unrelated to the current task, it still does not prevent energy consumption for sending raw images. One solution is to compress the images so the receiver can reconstruct and exploit its side knowledge. For example, one can deploy a generative model, such as a \gls{gan}, to compress the raw images, where recent studies~\cite{HiFiC,Tiny_HiFiC} show that it can achieve high perceptional quality compared with conventional image compression methods.

\begin{figure}[t]
    \centering
    \includegraphics[width=0.48\textwidth]{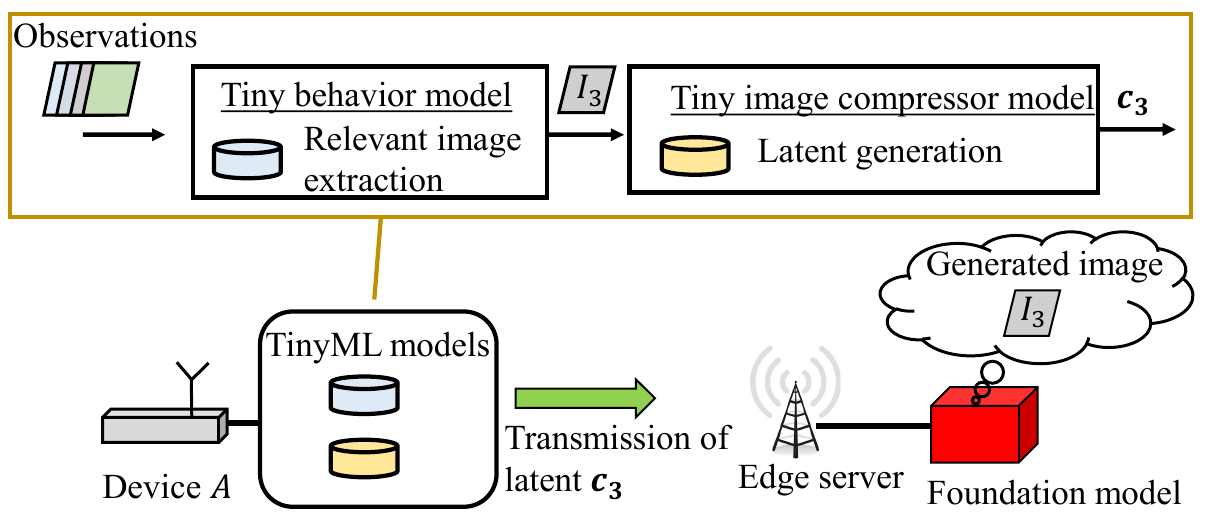}
    \caption{EcoPull, a sustainable \gls{iot} image retrieval framework in which two types of Tiny\gls{ml} models are installed into \gls{iot} devices: i) a \textit{behavior model} and ii) an \textit{image compressor model.}}
    \label{Fig:Toy_Example}
    \vspace{-4mm}
\end{figure}

\paragraph*{Contributions} In this work, we present EcoPull, an \gls{iot} image retrieval framework in which the \gls{es} pulls that from \gls{iot} devices equipped with two types of Tiny\gls{ml} models: i) a \textit{behavior model} and ii) an \textit{image compressor model}. Fig.~\ref{Fig:Toy_Example} exemplifies the proposed framework, where the \gls{es} equipped with a \gls{fm} attempts to obtain the information that an \gls{iot} device $A$ observed at a given time. Each \gls{iot} device first checks the relevance of the images against the \gls{es}'s current task by leveraging the installed behavior model to filter out irrelevant images, \emph{e.g.}, by conducting feature extraction and calculating a matching score as in~\cite{shiraishi2023energy,semdas}. Next, the \gls{iot} device encodes the relevant images, resulting in latent feature vector representations, which we refer to as \emph{latents}~\cite{ramesh2022hierarchical}. The \gls{es} reconstructs the latents using a decoder. We show that the interplay of these two models allows us to reduce the energy consumption of \gls{iot} devices to a greater extent while maintaining the quality of the retrieved images at the \gls{es}. In particular, we introduce a new performance metric called \textit{\gls{sifi} that jointly evaluates the significance and the fidelity of the image retrieved at the \gls{es}.} We describe our framework from a \gls{mac} layer perspective, focusing on the scenario where multiple devices share a common medium.

\paragraph*{Related work} We consider that our image compressor model is based on a generative model. \gls{gai} has been introduced in communication systems, such as in the network layer~\cite{thorsager2024generative} and the federated learning setup~\cite{huang2024federated}. In addition, the concept of \gls{giot} was introduced in \cite{wen2023generative}, in which a potential application and its challenges are discussed. In \cite{du2023generative}, the authors introduced \gls{gai}-aided semantic communication to realize accurate content decoding. Our work relates to these in the sense of only transmitting the intended meaning over the channel thanks to the compression and shared knowledge~\cite{barbarossa2023}. However, we focus on reducing and analyzing the overall energy consumption of \gls{iot} devices considering the additional energy cost of the \gls{ml} models, which has been overlooked in the literature.


\section{The EcoPull Framework}\label{sec:Framework}

\subsection{Scenario}\label{sec:preliminaries}
We consider a scenario where an \gls{es} is interested in retrieving images from $K$ \gls{iot} devices, such as mobile robots or fixed-mounted cameras, deployed over a sensing field to fulfill a task requested by a user, such as a mobile user, through the cloud via a \textit{query}. The \gls{es} has a \gls{fm} to \textit{interpret} the user's query. For example, the query could comprise a \texttt{prompt}, and the \gls{fm} could be based on a \gls{gpt} architecture. We then consider a \textit{pull-based communication system} where the \gls{es} retrieves data from the devices based on a piece of \textit{side information} given by the \gls{es}' \emph{interpretation} of the query sent by the user. The \gls{es} and \gls{iot} devices communicate over an idealized slotted multiaccess channel where \textit{packets}, which represent the minimal piece of information, have the same length of a \textit{slot} of duration $T_{p}$ [s] or equivalently $b_{p}$ bits~\cite{Bertsekas1996}. We define a \textit{frame} as a collection of $L$ \textit{slots} whose total number is $F$. We assume an error-free downlink channel and that the packets are lost in case a collision in the uplink occurs during a slot.

The $i-$th device, where $i \in \mathcal{K}=\{1,\dots,K\}$, locally stores the collected images in unlabeled datasets $\mathcal{D}^{i} = \{\mathbf{x}_1^{i},\ldots,\mathbf{x}^{i}_{N}\}$. The number of images is denoted by $|\mathcal{D}^{i}|=N^{i}$ and $\mathbf{x}^{i}_n\in\mathbb{R}^{M_{C}\times {M_H} \times M_{W}}$ represents the original image with $M_{C}$ channels, where $M_H$ and $M_W$ are the height and width of the image in pixels, respectively. When referring to a generic image, we drop the indices $i$ and $n$ as $\mathbf{x}$. The index set of images at the $i$-th device is denoted as $\mathcal{N}^{i}=\{1,\dots,N^{i}\}$. As in~\cite{moons2017minimum}, we assume each device has fixed-point arithmetic hardware equipped with a two-dimensional processing chip for \glspl{cnn}. This consists of an off-chip \gls{dram} that works with $b_{\max}$ quantization bits (full precision), a parallel neuron array with $p$ \glspl{muac} units of $b_{\text{MUAC}}$ bits, and two on-chip memory levels: a main \gls{sram} buffer that stores, \emph{e.g.}, the weights and activations, and a local \gls{sram} buffer that caches currently used weights and activations. We let $b_q$ be the number of bits for quantization in the \gls{sram}. Each unlabeled dataset $\mathcal{D}^i$ is stored in the \gls{dram} with precision $b_{\max}$. Moreover, each device has stored two Tiny\gls{ml} models in the \gls{dram}: i) a \emph{behavior model} and ii) an \emph{image compressor model}, see Sec.~\ref{sec:overview}.

\subsection{Description of the Scheme}\label{sec:overview}
The proposed scheme comprises four distinct phases:

\textit{\underline{Phase 1. Downlink pulling phase:}} Triggered by a query from a user, the \gls{es} fetches data from the devices. First, the \gls{es} transmits the trained Tiny\gls{ml} models towards the devices to facilitate the subsequent semantic query, as in~\cite{shiraishi2023energy} -- this can be done as the models are updated.\footnote{In particular, we assume that the behavior model is sent more frequently than the image compressor since its accuracy depends on the current query. Therefore, we model the energy spent to receive the behavior model in Sec.~\ref{sec:energy}, while the cost to receive the image compressor is ignored.} Then, the \gls{es} broadcasts a $b_q$-quantized vector semantic feature vector $\mathbf{q}\in\mathbb{R}^{M}$ obtained by the \gls{fm} to the devices over the downlink channel.

\textit{\underline{Phase 2. Behavioral phase:}} After receiving the semantic query, the devices use their behavior model to compute a set of similarity measures $\{s_{n}^{i}\}_{n=1}^{N^{i}}$ between their images in $\mathcal{D}^{i}$ and $\mathbf{q}$, where $s_{n}^{i}\in[0,1]$. Let $V_{\mathrm{th}}\in[0,1]$ be a \textit{relevance threshold}, which can be set by the \gls{es}. The $i-$th device extracts the set of relevant images $\mathcal{D}^{i}_{s}\subseteq\mathcal{D}^{i}$ as
\begin{equation}
    \mathcal{D}^{i}_{s} = \{\mathbf{x}^{i}_n|s_{n}^{i}\geq V_{\rm th},\forall n\in\mathcal{N}^{i}\}.
    \label{eq:relevant-images}
\end{equation}
Let $|\mathcal{D}^{i}_{s}|=S^{i}$ be the number of relevant images. {If $S^{i} > 1$, the $i-$th device is a \textit{relevant device} for the given query.} The set of relevant devices $\mathcal{K}_s\subseteq\mathcal{K}$ is defined as $\mathcal{K}_s = \{i|S_i>0,\forall i\in\mathcal{K}\}$, where $|\mathcal{K}_s|=K_s$ is the number of relevant devices.

\textit{\underline{Phase 3. Uplink compression-then-transmission phase:}} At the beginning of this phase, the relevant devices compress their relevant images using the image compressor model. Let $C(\cdot)$ denote the tiny encoder and $\mathbf{z}\in\mathbb{R}^{M^{\prime}_C \times M^{\prime}_H \times M^{\prime}_W}$ denote a quantized latent variable, whose quantitation occurs according to {$b_{\max}$} and where $M^{\prime}_C$ is the new number of channels and $M^{\prime}_H$ and $M^{\prime}_W$ are the new height and width of the compressed image. Thus, the encoding is represented as $\mathbf{z}=C(\mathbf{x})$. For the $i-$th device, the compressed set of relevant images is:
\begin{equation}
    \check{\mathcal{D}}^{i}_{s} = \{\mathbf{z}^{i}_n|\forall \mathbf{x}^{i}_n\in{\mathcal{D}}^{i}_{s}\}, 
    \label{eq:compression}
\end{equation}
with $|\check{\mathcal{D}}^{i}_{s}|=S^{i}$. Each device stores the $S^{i}$ images in a queue $\mathcal{Q}^{i}\gets\check{\mathcal{D}}^{i}_{s}$ in the \gls{dram} with precision $b_{\max}$. Next, the relevant devices transmit their compressed relevant images over the $f-$th frame, where we tie the size of a packet to the size of the latent as $b_p=M^{\prime}_C \cdot M^{\prime}_H \cdot M^{\prime}_W\cdot b_{\max}$ bits, where we ignore the header size for simplicity of analysis.

For simplicity, we assume that each relevant device attempts to transmit a single image per frame by choosing a slot at random. With this approach, a number of $K_s$ or less devices compete for the $L$ slots at a given frame. We assume they do so by adopting the slotted ALOHA protocol~\cite{Bertsekas1996} without re-transmissions to showcase the main benefits of our framework and ease the analysis. After an attempt to transmit an image, the queue is updated as $\mathcal{Q}^{i}\gets\mathcal{Q}^{i}\setminus{\mathbf{z}^{i}_{f}}$, where $\mathbf{z}^{i}_{f}$ denotes the image selected at random by the $i-$th device to be transmitted at the $f-$th frame. When a device empties its queue, it becomes idle and stops competing in the next slots.

\textit{\underline{Phase 4. Data decompression and response to the user:}} Let $G(\cdot)$ be the tiny decoder stored at the \gls{es}. After the $F$ frames are over, the \gls{es} collects and decompresses all the non-colliding packets, whose decoding operation is denoted as $\hat{\mathbf{x}}=G(\mathbf{z})$. Then, the \gls{es} identifies the top $l$ most relevant images for the specific query, with, \emph{e.g.}, $l=3$, to be sent as a reply to the user as a form of text or image by exploiting the \gls{fm}. To measure the system's performance, we introduce \gls{sifi} in Sec.~\ref{sec:performance}.

Based on~\cite{shiraishi2023energy}, we assume that the behavior model is a multimodal feature extractor $\zeta:\mathbf{x}\mapsto\mathbf{o}$ based on a \gls{fm} that can take as input images/texts $\mathbf{x}$ and has as output $\mathbf{o}\in\mathbb{R}^{M}$. The \gls{es} should compress $\zeta$ -- \emph{e.g.}, by using pruning -- to send it to the devices, as described in \underline{Phase 1}. Because of that, we assume that the devices receive a model $\tilde{\zeta}$ that performs worse than ${\zeta}$. Given that and the description of \underline{Phase 2}, the similarity measure at the $i-$th device is computed as $s^{i}_n=g(\tilde{\zeta}(\mathbf{x}^{i}_n),\mathbf{q})$, where $g(\cdot)$ is a similarity function, such as cosine similarity.

The fact that $\tilde{\zeta}$ performs worse than $\zeta$ is modeled by how much the similarity measure $s^{i}_n$ deviates from the true similarity measure $\beta^{i}_n=g({\zeta}(\mathbf{x}^{i}_n),\mathbf{q})$. Specifically:
\begin{equation}
    s^{i}_n = \beta^{i}_n + w^{i}_n,
    \label{eq:error-behavior-model}
\end{equation}
where $w^{i}_n  \sim \mathcal{N}(0, \sigma_{\mathrm{ML}}^2)$ is Gaussian distributed and $\sigma_{\mathrm{ML}}$ is a standard deviation representing the model noise. For simplicity, we model the quantization effect from getting $\tilde{\zeta}$ of $\zeta$ by setting $\sigma_{\mathrm{ML}}= \frac{1}{b_{B}}$, where ${b_{B}}\leq b_{\max}$ is the number of bits for the quantization of weights of the behavior model for transmission~\cite{shiraishi2023energy}. 

As for the tiny image compressor model, we adopt the framework presented in~\cite{Tiny_HiFiC} that introduces a generative image compression method. Here, we assume that $C(\cdot)$ and $G(\cdot)$ are jointly trained before the model deployment, \textit{e.g.}, at the cloud. We refer the interested reader to~\cite{Tiny_HiFiC} for more details about training and architecture. The compression rate $r$, measured in \gls{bpp}, is
\begin{equation}
    r = \dfrac{M_C^{\prime} \cdot M^{\prime}_H \cdot M^{\prime}_W \cdot b_{\max}}{M_H \cdot M_W}.
    \label{eq:compression-rate}
\end{equation}

\section{Energy Consumption}\label{sec:energy}
We consider the energy cost of communication and computation at the \gls{iot} devices while ignoring the energy consumed during idle periods. We let $\xi_{T}$~[W] be the energy cost per uplink transmission and $\xi_{R}$~[W] be the energy cost for receiving. Next, a general description of the \textit{energy cost per inference of a model}, $E_{\mathrm{inf}}$, follows based on~\cite{moons2017minimum}.\footnote{We assume that the main \gls{sram} of the devices is large enough to store feature maps and to hold the entire Tiny\gls{ml} model after being loaded from the off-chip \gls{dram}; unused feature maps are erased from the memory.} The energy cost per inference of a model is given by:
\begin{equation}
    E_{\mathrm{inf}} = E_{\mathrm{DRAM}} + E_{\mathrm{HW}},
    \label{eq:inference}
\end{equation}
comprised of the energy cost of accessing the \gls{dram} where the model and input are stored, denoted as $E_{\mathrm{DRAM}}$, and that of the actual processing for inference denoted as $E_{\mathrm{HW}}$. Then, $E_{\mathrm{DRAM}}$ can be expressed as follows:
\begin{equation}    
    E_{\mathrm{DRAM}}=E_{D} \cdot [M_{C} \cdot M_H \cdot M_W] \cdot (b_{\max} / b_q),  
    \label{eq:dram}
\end{equation}
where $E_{D}$ is the energy consumed per $b_q$-bit \gls{dram} access and $M_{C}\cdot M_H \cdot M_W$ is the input size. While $E_{\mathrm{HW}}$ can be modeled as the sum of the computing energy, $E_{C}$, and the costs of moving weights and activations from the main to the local \gls{sram}, $E_{W}$ and $E_{A}$, respectively, as described below: 
\begin{equation}
    E_{\mathrm{HW}} = E_{C} + E_{W} + E_{A},
    \label{eq:HW}
\end{equation}
with $E_{C} = E_{\mathrm{MUAC}} (U+3 A)$, $E_{W}= E_{M} W + E_{L} U/ \sqrt{p}$, and $E_{A} = 2 E_{M} A + E_{L} U/\sqrt{p}$, where $E_{\mathrm{MUAC}}$ is the energy consumed for a single \gls{muac} operation, $E_{L}=E_{\mathrm{MUAC}}$ is the energy consumption to read/write from/to the local \gls{sram}, and $E_{M} = 2  E_{\mathrm{MUAC}}$ is the energy consumed for accessing the main \gls{sram}. Other parameters are $U$, which denotes the network complexity in the number of \gls{muac} operations; $W$, which denotes the model size in the number of weights and biases; $A$, which denotes the total number of activations throughout the whole network. 

We now instantiate the inference model in~\eqref{eq:inference} for the behavior and the image compressor models. For the behavior model, we let $E_{\mathrm{inf}, B}$ denote the energy cost per inference and $(U_{B},W_{B},A_{B})$ be the set of parameters. For the image compressor model, we let $E_{\mathrm{inf}, C}$ denote the energy cost per inference and $(U_{C},W_{C},A_{C})$ be the set of parameters. 

For the $i-$th device, the \textit{total energy consumed for computation} can be written as:
\begin{equation}
    E^{i}_{\mathrm{comp}} = N^{i} E_{\mathrm{inf}, B} + S^{i} E_{\mathrm{inf}, C} + E_{D} \cdot ( W_B + W_C) \cdot (b_{\max} / b_q),
    \label{eq:energy-comp}
\end{equation}
the last term of sum measures the cost of loading the models from the \gls{dram} to the main \gls{sram}. While the \textit{total energy consumed for communication} can be written as:
\begin{equation}
    E^{i}_{\mathrm{comm}} = \dfrac{(\xi_R \cdot (W_B b_{B} + M b_{q})  + \xi_T \cdot S^{i} b_{p})}{R}, 
\end{equation}
where $\xi_R$ multiplies the time of receiving the behavior model and the vector $\mathbf{q}$, $\xi_R$ multiplies the time of transmitting the $S^{i}$ relevant images, and $R$ is the rate of the downlink and uplink channels in bits per second. \textit{The total energy consumed per device} is then:
\begin{equation}
    E^{i}_{\mathrm{tot}} =  E^{i}_{\mathrm{comp}} + E^{i}_{\mathrm{comm}}.
\end{equation}
We note that $S^i$ is a random variable that depends on $V_{\mathrm{th}}$ as per~\eqref{eq:relevant-images}, so is $E^{i}_{\mathrm{tot}}$. Let $P_{\mathrm{th}} (V_{\mathrm{th}})$ denote the probability that an image is relevant, \textit{i.e.}, the probability $s_{n}^i$ in eq.~\eqref{eq:error-behavior-model} is equal to or higher than $V_{\mathrm{th}}$. We can then derive the \textit{expected total energy consumed} given the model in~\eqref{eq:error-behavior-model}. This can be expressed by considering the distribution of observed similarity measure $s$ given true similarity measure $\beta$, $p(s|\beta) = \frac{1}{\sqrt{2\pi}\sigma^2}\int_{-\infty}^{\infty}\exp{\left(-\frac{(s-\beta)^2}{2\sigma_{\mathrm{ML}}^2}\right)}ds$, as follows:
 \begin{equation}
     P_{\mathrm{th}}\left(V_{\mathrm{th}}\right) = \int_{0}^{1}\,Q\left(\frac{V_{\mathrm{th}}\,-\,\beta}{\sigma_{\mathrm{ML}}}\right)\,g_{T}(\beta)\,d{\beta},
 \end{equation}
where $Q(x)$ denotes the Q-function defined as $Q(x) = \frac{1}{\sqrt{2\pi}}\int_{x}^{\infty}\,\exp\left(-\frac{u^2}{2}\right)du$ and $g_{T}(\beta)$ is the distribution of true similarity measure $\beta$. We let $B(p,n,k)$ be the binomial distribution, with probability $p$ and total number of elements $n$ with $B(p,n,k) = \binom{n}{k}p^{k}(1-p)^{n-k}$. Then, the probability that a device has $\nu$ relevant images out of $N^{i}$ images can be expressed as $P_{\mathrm{Rel}}(\nu) = B(P_{\mathrm{th}}\left(V_{\mathrm{th}}\right), N^{i}, \nu)$.
Finally, the \textit{expected total energy consumed per device} can be expressed as follows:
\begin{equation}
    \mathbb{E}[E_{\mathrm{tot}}^{i}]=\sum_{\nu=0}^{N^{i}}{\nu \left(E_{\mathrm{inf}, C}+\frac{{\xi_T}{b_{p}}}{R}\right)}P_{\mathrm{Rel}}(\nu) +\epsilon,
\end{equation}
where $\epsilon = N^{i} E_{\mathrm{inf}, B}+E_{D} \cdot ( W_B + W_C) \cdot (b_{\max} / b_q)+\xi_R (W_B b_{B} + M b_{q})/R$.

\section{Performance Analysis}\label{sec:performance}

\subsection{Measuring the significance and fidelity of retrieved images}\label{sec:utility}
To measure the performance of the retrieved image at \underline{Phase 4}, we introduce \gls{sifi}, a metric that can track two distinct objectives simultaneously: \textit{significance} and \textit{fidelity}. On the one hand, the significance represents how many \textit{actual relevant images} can be successfully collected, where the image whose true similarity measure $\beta$ is equal to or higher than $\delta$ is referred to as an \textit{actual relevant image}. On the other hand, the fidelity represents how similar the reconstructed image is to the original one. The fidelity can be measured over the reconstructed image, denoted as $k_{d}(\mathbf{x},\hat{\mathbf{x}})\in[0, 1]$, where $k_d$ is a perceptual quality measure~\cite{heusel2017gans}, such as \gls{fid}~\cite{heusel2017gans}, which evaluates to 0 when the reconstructed image is the same compared to the one observed by the \gls{iot} device. We note that $k_{d}(\mathbf{x},\hat{\mathbf{x}})$ varies according to the compression rate $r$, whose dependency can be denoted as $k_{d}(\mathbf{x},\hat{\mathbf{x}};r)$, as found in~\cite{thorsager2024generative}. {Finally, we can define the \gls{sifi} of the proposed framework, $\vartheta \in [0, 1]$. Let $T^{i}_n\in\{0,1\}$ be a Bernoulli random variable which models if the transmission of the $n$-th image of the $i$-th device was successful, and $\Omega$ be the subset of index pairs $(i,n)$ for the actual relevant images, \textit{i.e.}, $\Omega = \{(i, n)~|~\beta^{i}_n \geq \delta,\forall n\in\mathcal{N}^{i}, \forall i \in \mathcal{K}\}$ and $|\Omega|$ its cardinality. Then, the \gls{sifi} can be defined as:}
\begin{equation}
    \vartheta =
    \begin{cases}
        1-\frac{1}{|\Omega|}\sum_{(i, n) \in~\Omega}k_{d}(\mathbf{x}_n^i, \hat{\mathbf{x}}_n^i;r)^{T^{i}_n}\Gamma^{1-T^{i}_n},&\text{if}~|\Omega| \geq 1,\\ 
        1, & \text{otherwise},
    \end{cases}
    \label{eq:sifi}
\end{equation}
where $\Gamma\in[0,1]$ is a penalization term if the data transmission of an actual relevant image has failed. Here, $\Gamma$ should be chosen to be greater than $k_{d}(\mathbf{x}_n^i, \hat{\mathbf{x}}_n^i;r)$. {A higher \gls{sifi}, $\vartheta$, implies that the \gls{es} retrieves the useful data that represents the current/accurate views of the observations of \gls{iot} devices regarding the user's query.} \gls{sifi} can be improved by collecting more data at the expense of increased energy consumption.

\subsection{Expected SiFi}\label{sec:analysis_SiFi}
We derive the \textit{proposed framework's expected \gls{sifi}}. Without loss of generality, we assume that $N^{i} = N,\,\forall i\in\mathcal{K}$. We start by deriving the average transmission success probability, defined as the probability that a packet is delivered to the \gls{es} without collision. This probability depends on the number of relevant images each device has, $S^{i}$, which we model as follows. Let $\{X_{\nu}\}_{\nu = 0}^N$ denote the random variables representing the number of devices having $\nu$ relevant images. Then, $\mathbf{X}=\{X_{0}, \ldots, X_{N}\}$ follows a multinomial distribution, whose \gls{pmf} of $\mathbf{\psi}= \{q_{0}, \ldots, q_{N}\}$ is:
\begin{equation}
    \ \mathbb{P}(\mathbf{X} = \mathbf{\psi})=\begin{cases}K!\prod_{\nu=0}^{N}\frac{P_{\mathrm{Rel}(\nu)}^{q_{\nu}}}{q_{\nu}!},& ({\sum_{\nu=0}^{N}{q_{\nu}}} = K), \\0,& \text{otherwise}.\label{eq:JointDistribution}
\end{cases}
\end{equation}
Given a realization sample $\psi_{u} = \{q_{0}^{u}, q_{1}^u, \ldots, q_{N}^u\}$, the number of devices which attempts data transmission in the $f$-th frame can be expressed as follows:
\begin{equation}
    W_{f}^u=\sum_{j=f}^{N}q_{j}^{u}.\label{eq:available_packet}
\end{equation}
Using eq.~\eqref{eq:available_packet}, the total number of frames required to collect all relevant images, denoted as $n_{w}^u$, can be described as:
\begin{equation}
    n_{w}^u= \argmin_{J \in [0, N]}~\left(q_{0} + \sum_{f=1}^{J}q_{f}^u = K\right).
    \label{eq:n_w_u}
\end{equation}
This can be related to the number of frames $F$ that the \gls{es} sets. Ideally, we would like to have $F\leq n_{w}^u$. Based on the eqs.~\eqref{eq:available_packet} and \eqref{eq:n_w_u}, we can derive the average transmission success probability per relevant image as:
\begin{equation}
    \ P_{s}(\psi_{u})=\begin{cases}\frac{1}{n_{w}^u}\sum_{f=1}^{n_{w}^u}\left(1-\frac{1}{L}\right)^{W_{f}^u-1},& \text{if}~n_{w} \geq 1,\\1,&\text{otherwise}.\end{cases}
    \label{eq:success_probability}
\end{equation}
Then, we must consider how many relevant images succeed in data transmission. Recall from Sec.~\ref{sec:energy} that the actual relevant image is defined as the image whose true similarity measure $\beta$ is higher than the threshold $\delta$. Here, the probability of an image is actually relevant, denoted as $P_{\delta}$, can be expressed as $P_{\delta}=\int_{\delta}^{1}g_{T}(\beta)d{\beta}$. For the actual relevant data to be successfully delivered to the \gls{es} given $\psi_{u}$, it needs to satisfy two conditions: 1) the observed similarity measure $s$ in eq.~\eqref{eq:error-behavior-model} should be higher than $V_{\mathrm{th}}$, and 2) the transmission of the data should be successful. Accordingly, the probability that the \gls{es} can successfully collect an actual relevant image can be expressed with the conditional probability:
\begin{equation}
    \begin{split}
        P_A(\psi_{u})&=\frac{P_{s}(\psi_{u})}{P_{\delta}}\int_{\delta}^{1}{Q\left(\frac{V_{\mathrm{th}}-\beta}{\sigma_{\mathrm{ML}}}\right)}g_{T}(\beta)d{\beta}.
        \label{eq:pr_A}
\end{split}
\end{equation}

Now we can derive the expected \gls{sifi} in eq.~\eqref{eq:sifi}. First, we focus on the case where $|\Omega|~\geq~1$, which happens with the probability $P_{\Omega} = 1- (1- P_{\delta})^{KN}$. Let $\kappa > 0$ denote the random variable representing the total number of actual relevant images in the sensing field and $Z_{\iota}^u$ be the random variable for $\iota$-th actual relevant image given $\phi_{u}$, which takes the value of $1-k_{d}(\mathbf{x}_n^i, \hat{\mathbf{x}}_n^i|r)$ if its image is delivered to the \gls{es} without collisions; otherwise, $\Gamma$. Then, its expectation is:
\begin{equation}
    \mathbb{E}[Z_{\iota}^{u}]= (1-k_{d}(\mathbf{x}_n^i, \hat{\mathbf{x}}_n^i|r))P_A(\psi_{u})+(1-P_A(\psi_{u}))(1-\Gamma). 
\end{equation}
Using this, the expected \gls{sifi} for $|\Omega| \geq 1$, given $\psi_{u}$ can be calculated as $\frac{1}{\kappa}\mathbb{E}[\sum_{\iota=1}^{\kappa}Z_{\iota}^u] = \frac{1}{\kappa}\sum_{\iota=1}^{\kappa}\mathbb{E}[Z_{\iota}^u] = \mathbb{E}[Z_{\iota}^u]$, where we apply the linearity of expectation for the independent random variables $Z_{1}^u, Z_{2}^u, \ldots, Z_{\kappa}^u$. On the other hand, the case of $|\Omega| = 0$ happens with probability $1 - P_{\Omega}$, in which case the \gls{sifi} is 1, as in eq.~\eqref{eq:sifi}. Finally, from the above description, the expected value of \gls{sifi} can be expressed as follows:
\begin{equation}
    \mathcal{V}=\sum_{u=1}^{|\psi|}\mathbb{P}(\psi_{u})\left(\mathbb{E}[Z_{\iota}^u]P_{\Omega}+(1- P_{\Omega})\right).
    \label{eq:expected-sifi}
\end{equation}

\subsection{Getting approximation of expected \gls{sifi}}
As the calculation of \gls{sifi} in eq.~\eqref{eq:expected-sifi} requires consideration of all $\tbinom{K+N}{K}$ realizations of multinomial distribution given in eq.~\eqref{eq:JointDistribution}, we resort to the \gls{mcmc} method~\cite{Bishop} to obtain approximate results. Specifically, we employ the Metropolis algorithm~\cite{Bishop} that generates a sequence of samples (realizations) $Y^{(t)}~ (t \in [0, T])$ based on probabilistic rules. Here, we denote the number of devices that have $\nu$ relevant images at the $t$-th realization sample as $q_{\nu}^{(t)}$, which needs to satisfy the constraint, $\sum_{\nu=0}^{N}{q_{\nu}^{(t)}}=K$ for all $t$. Here, an initial state $Y^{(0)}$ is set to fairly distribute the number of devices observing $\nu$ images. Then, the sequence of samples is generated as follows:

\textit{\underline{Step~1. Generate new sample:}}~Create a new sample $Y^{'}$ from the current state $Y^{(t)}$. First, we randomly select one value $a ~\in~[0, N]$ out of $N + 1$ values, satisfying $q_{a}^{(t)}~{\geq}~1$ and decrease $q_a^{(t)}$ by 1. Then, we randomly select another value $b$ satisfying $q_{b}^{(t)}~{<}~N$ and increase $q_b^{(t)}$ by 1. With these operations, a new sample $Y^{'}$ can be generated.

\textit{\underline{Step~2. Calculate transition cost:}}~Calculate transition cost as $\mathbb{P}(Y^{'})/{\mathbb{P}(Y^{(t)})}$, by using \gls{pmf} given by eq.~\eqref{eq:JointDistribution}. More details on transition cost can be found in~\cite{Bishop}.

\textit{\underline{Step~3. Update state:}}~Generate a random number $r'~{\in}~[0,1]$, following uniform distribution, and decide the next state as: 
\begin{equation}
    Y^{(t+1)}=
    \begin{cases}
        Y^{'},&(\text{if}~r'~{\leq}~\frac{\mathbb{P}(Y^{'})}{\mathbb{P}(Y^{(t)})}),\\Y^{(t)},&\text{otherwise}.
    \end{cases}
\end{equation}
The above operation is repeated over $T$ times, and for each $Y^{(t)}$, the \gls{sifi} is calculated through the equations derived in Sec.~\ref{sec:analysis_SiFi} and stored as $u^{(z)}$. Finally, the approximate \gls{sifi} is:
\begin{equation}
    \hat{\mathcal{V}}=\frac{1}{T}{\sum_{t=1}^{T}{u^{(t)}}}.
    \label{eq:MCMCdelay}
\end{equation}

\section{Numerical Evaluation}
This section investigates the performance of EcoPull. We consider images as ($M_{C}$, $M_{H}$, $M_{W}$) = (3, 640, 480)~\cite{Tiny_HiFiC}. The values of $E_{\mathrm{MUAC}}$ and $E_{D}$ are: $E_{\mathrm{MUAC}}= 3.7\, \cdot ({b_{q}}/{b_{\mathrm{MUAC}}})^{1.25}$ [pJ] and $E_{D}= 128 \cdot 3.7 \cdot ({b_{q}}/{b_{\mathrm{MUAC}}})$ [pJ], respectively~\cite{moons2017minimum}. Moreover, we use $b_{\mathrm{max}} = 16$, $b_{\mathrm{MUAC}} = 16$, $b_{B} = 8$, $b_{q} = 8$, and $p = 64 \cdot {b_{\mathrm{MUAC}}}/{b_{q}}$ for the behavior model based on~\cite{moons2017minimum} and $b_{q} = 16$ for the image compressor model~\cite{Tiny_HiFiC}. Using~\cite{vazquez2015modeling}, we set $\xi_{T} = 108$ [mW], $\xi_{R} = 66.9$ [mW]. For the behavior model, ($U_B$, $W_B$, $A_B$) = (117 M, 0.976 M, 4.309 M)~\cite{xu2022etinynet}. For the image compressor model, ($U_C$, $W_C$, $A_C$) = (477 M, 0.0184 M, 3.54 M)~\cite{Tiny_HiFiC}. We set $R = 100$ [kbps], $\delta = 0.9$, and number of \gls{mcmc} rounds $T = 10^4$. For simplicity, we generate $g_{T}(\beta)$ based on uniform distribution within $[0,1]$ and assume that $N^{i} = N,\,\forall i\in\mathcal{K}$. We conduct the Monte Carlo simulation over $10^4$ rounds to obtain the averaged results. 

{The values of $k_{d}(\mathbf{x},\hat{\mathbf{x}})$ in eq.~\eqref{eq:sifi} is set as $k_{d}(\mathbf{x}_n^i, \hat{\mathbf{x}}_n^i|r)$ = $0.0725^{r}$ based on the results of our curve fitting~\cite{thorsager2024generative} in terms of normalized \gls{fid} using \gls{hific}, in which images are taken from the COCO2017~\cite{lin2014microsoft}, {and we set $\Gamma = 1$}.} We argue these results are representative of the performance of the tiny image compressor in~\cite{Tiny_HiFiC} based on the \gls{fid} results for the nonquantized network, which is shown to be similar in performance to the original image compressor of \gls{hific}~\cite{HiFiC}. 

As benchmarks, we use the conventional compression method referred to as \textit{baseline} and \textit{TinyAirNet}~\cite{shiraishi2023energy}. For the baseline, each device first compresses its image with PNG and transmits it in predetermined slots \textit{without collision}. For the {TinyAirNet}, each device only uses a behavior model and transmits PNG-compressed images. For brevity, we omit the analysis of both schemes.

\begin{figure}[t]
    \centering
    \input{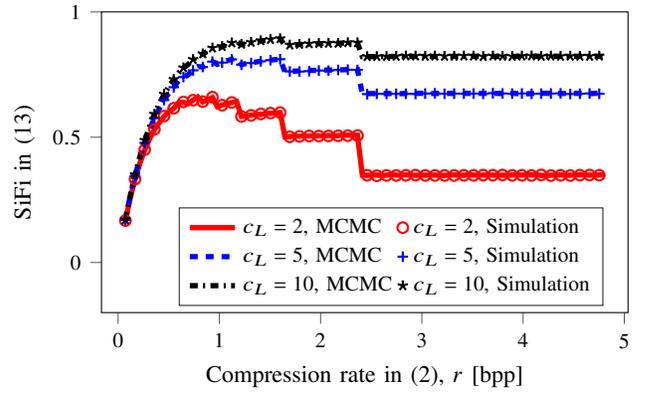}
    \caption{\gls{sifi} in \eqref{eq:sifi} against the compression rate $r$ in [bpp]. `MCMC' refers to the approximation of the expected \gls{sifi} in~\eqref{eq:expected-sifi}, while `Simulation' refers to the one obtained by averaging~\eqref{eq:sifi}. The number of available transmission slots $L$ is calculated as $L = {c_{L}} \lceil r^{\mathrm{max}}/{r} \rceil$, where $r^{\mathrm{max}}=4.86$ is the average \gls{bpp} for the PNG compressed images and $c_{L}$ is a coefficient value~\cite{thorsager2024generative}.}
    \label{Fig:utility_against_Bp}
    \vspace{-4mm}
\end{figure}

Fig.~\ref{Fig:utility_against_Bp} shows the \gls{sifi} against the compression rate $r$, where we set $V_{\mathrm{th}} = 0.6$, $K = 5$, $c_{L} \in \{2, 5, 10\}$ and $N = 100$. From this figure, first, we can see that the results for EcoPull obtained by our approximate analysis using \gls{mcmc} coincide with that of simulation results, which validates our approach using \gls{mcmc}. In the figure, we can see the saw-tooth pattern because of the discrete nature of available transmission slots $L$ for each $r$, \textit{e.g.}, for $r \in [1.204, 1.608]$, \gls{sifi} slightly increases against $r$, in which $L$ is a constant. Next, the figure also shows an optimal compression rate in terms of the \gls{sifi}. Let us denote the optimal value of $r$ as $r^{\mathrm{opt}}$. For $r^{\mathrm{opt}} > r$, the transmission success probability becomes larger thanks to the lower compression rate, but the \gls{fid} is high, leading to the poor \gls{sifi}. For $r^{\mathrm{opt}} < r$, the \gls{fid} of the retrieved image is low, but the transmission success probability of the relevant image becomes small because of the smaller number of available transmission slots. Further, the \gls{sifi} becomes larger as $c_{L}$ increases. Each device with the relevant image can choose the transmission slot from the larger number of available slots $L$, increasing the transmission success probability. Finally, from this result, we can see the importance of the choice of the compression rate in accordance with the available transmission slots to maximize the \gls{sifi}. 

The total energy consumption and \gls{sifi} of EcoPull depends on the value of $V_{\mathrm{th}}$, as well as $r$, as mentioned above. For example, the smaller (higher) $V_{\mathrm{th}}$ increases (decreases) the \gls{sifi}, but total energy consumption becomes larger (smaller). To realize a fair comparison among the different schemes, we investigate the minimum energy consumption under the constraint of a \gls{sifi} value $\gamma_{\mathrm{th}}$. To this end, we first obtain the subsets of parameters $\{V_{\mathrm{th}}, r\}$ for EcoPull, denoted as $\mathcal{P}(N)$ satisfying the constraint for given the value of $N$:
\begin{equation}
    {\mathcal{P}(N)=\min_{V_{\mathrm{th}}, r} \mathbb{E}[E_{\mathrm{tot}}^i] (N)~~\text{s.t.}~\hat{\mathcal{V}} \geq \gamma_{\mathrm{th}}.}
\end{equation}
We solved this numerically through a grid search, varying the value of $V_\mathrm{th}$ within a range of [0.50, 0.80] with a step of 0.01 and that of $r$ within a range of [1, 2] with a step of 0.667.

\begin{figure}[t]
    \centering
\begin{tikzpicture}
\definecolor{chocolate2267451}{RGB}{226,74,51}
\definecolor{dimgray85}{RGB}{85,85,85}
\definecolor{gainsboro229}{RGB}{229,229,229}
\definecolor{steelblue52138189}{RGB}{52,138,189}

\begin{axis}[
width=7cm,
height=4cm,
scale only axis,
xlabel={Total number of images per device, $N$},
xmin=0, xmax=100,
xtick style={color=dimgray85},
ylabel={Energy saving w.r.t. baseline, $\eta$},
ymin=0, ymax=4.5,
ytick style={color=dimgray85}
]

\addplot [thick, black, dashed, line width = 2]
table
{
x  y
-5 1
105 1
};
\addlegendentry{\text{Baseline scheme}}

\addplot [thick, dashdotted, blue, line width=2]
table {%
5 3.98613454100712
10 2.07328879021673
15 1.42434978116949
20 1.1150789118289
25 0.930084350445263
30 0.808275350431835
35 0.726323045030185
40 0.658099776458199
45 0.610287541250733
50 0.577497836568901
55 0.543437364184051
60 0.511616814582757
65 0.492903469723815
70 0.473354986867819
75 0.457600278739384
80 0.441570956466333
85 0.429739525607329
90 0.418785813227474
95 0.405796083693892
100 0.397780817950862
};
\addlegendentry{\text{TinyAirNet}~\cite{shiraishi2023energy}}

\addplot [thick, red,line width=2]
table {%
5 3.60816800152869
10 1.82083374665893
15 1.2227070362874
20 0.926795982518943
25 0.749367149111988
30 0.631396665379047
35 0.548180294032316
40 0.484366144261834
45 0.435821866429834
50 0.398118903013229
55 0.365770946959082
60 0.338101496277999
65 0.316392054571194
70 0.297056280265856
75 0.280544865529105
80 0.268894686077017
85 0.256231489307219
90 0.244878280457279
95 0.234011847437143
100 0.2250483993448
};
\addlegendentry{\text{EcoPull}}

\end{axis}

\end{tikzpicture}
    \caption{Energy saving w.r.t. the baseline $\eta$ as a function of the total number of images per device, $N$.}
    \label{Fig:Ene_ratio}
    \vspace{-4mm}
\end{figure}
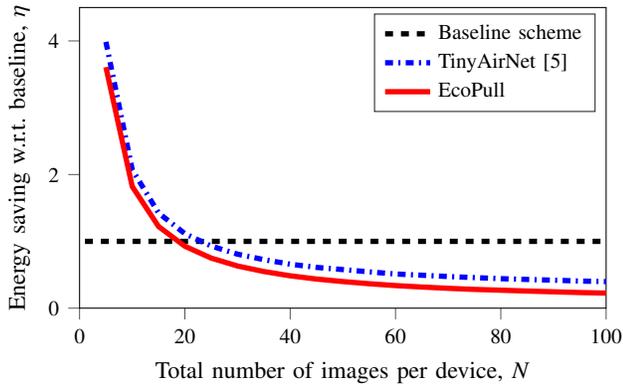

Fig.~\ref{Fig:Ene_ratio} shows the energy saving w.r.t. the baseline $\eta$ as a function of the number of images, $N$, where we set $K = 5$, $c_{L} = 5$, and $\gamma_{\mathrm{th}} = 0.8$. Here, $\eta$ is defined as the ratio between the total amount of energy consumed by the given scheme compared with the baseline scheme. Note that $\eta$ becomes smaller than 1 only if the total energy consumption of the given method is smaller than the baseline scheme. The figure shows that $\eta$ of EcoPull and TinyAirNet becomes smaller as $N$ becomes larger. This is because as $N$ increases, the gain from filtering out irrelevant images becomes larger than the additional energy cost brought by the Tiny\gls{ml} model. Finally, we can see that EcoPull can realize a smaller $\eta$ than no latent transmission scheme while satisfying the utility requirements because of the reduction of transmission data size for the relevant images. These results indicate that our proposed scheme can realize high energy efficiency while maintaining higher \gls{sifi}, especially when the number of images is high.

\section{Conclusions}
In this paper, we have proposed EcoPull, an \gls{iot} data collection framework empowered by TinyML models, which aims to reduce the total energy consumed by \gls{iot} devices. We have introduced two types of Tiny\gls{ml} models to \gls{iot} devices: the behavior model and the image compressor model. In our framework, each device first suppresses irrelevant data transmission using the behavior model and then transmits compressed images by applying an image compressor model. We have derived equations expressing total energy consumption and the \gls{sifi}, to evaluate the proposed framework. Numerical results have revealed that our proposed framework achieves high energy efficiency while maintaining high retrieved performance, especially when the number of images is large.

\bibliographystyle{IEEEtran}
\bibliography{IEEEabrv,Ref}

\end{document}